\documentclass[twocolumn,showpacs,superscriptaddress,preprintnumbers,amsmath,amssymb,prc]{revtex4-1}

\usepackage{graphicx}
\usepackage{dcolumn}
\usepackage{bm}
\usepackage{hyperref}
\usepackage{soul} 
\usepackage{manfnt} 
\usepackage{url}

\soulregister\cite7
\soulregister\ref7

\begin{document}

\title{Synthesis of the superheavy elements beyond Og: extrapolating from ${}^{48}\mathrm{Ca}$ to ${}^{54}\mathrm{Cr}$}

 \author{Yueping Fang}
 \affiliation{Sino-French Institute of Nuclear Engineering and Technology, Sun Yat-sen University, Zhuhai 519082, China}




 \author{Long Zhu}
 \email[Corresponding author, ]{zhulong@mail.sysu.edu.cn}
 \affiliation{Sino-French Institute of Nuclear Engineering and Technology, Sun Yat-sen University, Zhuhai 519082, China}
\affiliation{Guangxi Key Laboratory of Nuclear Physics and Nuclear Technology, Guangxi Normal University, Guilin 541004, China}


\date{\today}

\begin{abstract}

Theoretical predictions on the optimal reaction energies are essential for producing superheavy elements (SHEs) beyond Og. Due to the limitation of the targets, synthesizing elements 119 and 120 will require beams of ${}^{50}\mathrm{Ti}$ and/or ${}^{54}\mathrm{Cr}$ ions. However, is it reliable to theoretically extrapolate from the well-investigated ${}^{48}\mathrm{Ca}$ induced reactions to those with heavier projectiles? In this work, we apply the Fusion-by-Diffusion (FBD) concept to answer this question from two perspectives: radial and mass asymmetry degrees of freedom. The FBD concept is employed in the mass asymmetry degree of freedom for the first time, 
in which by fitting the calculations to experimental evaporation residue cross sections (ERCS) for the reactions of ${}^{48}\mathrm{Ca}$ as projectiles with the actinide targets, a strong linear correlation between the contact distance ($D_\mathrm{cont}$) and center-of-mass energy excess  above the Coulomb barrier
($E_\mathrm{c.m.}-B_0$) is found and a parametrization of the $D_\mathrm{cont}$ is introduced. Using the result of parametrization, the calculations satisfactorily reproduce the shapes of all hot fusion excitation functions and values of the ERCS. Furthermore, thanks to the recent experimental data, we extrapolate the calculation in the reactions \({}^{50}\mathrm{Ti}+{}^{242}\mathrm{Pu}\), \({}^{50}\mathrm{Ti}+{}^{244}\mathrm{Pu}\), and \({}^{54}\mathrm{Cr}+{}^{238}\mathrm{U}\). The calculations reproduce the experimental data rather well within the experimental errors in both perspectives. Our results demonstrate that there is no non-negligible systematic deviation in extrapolating the projectiles from $^{48}\text{Ca}$ to $^{50}\text{Ti}$ and $^{54}\text{Cr}$ for synthesizing SHEs beyond Og.

\end{abstract}
 
\maketitle


$\emph{Introduction.}$ The synthesis of superheavy elements (SHEs) is a frontier of research in nuclear physics 
\cite{oganessianSuperheavyElementResearch2015a,nazarewiczLimitsNuclearMass2018,giulianiColloquiumSuperheavyElements2019,hofmannDiscoveryHeaviestElements2000,gates2024toward}. With regard to the hot fusion reactions, a remarkable progress has been made in the synthesis of SHEs by employing double magic projectile \(^{48}\text{Ca}\) and actinide targets \cite{oganessianSynthesisNucleiSuperheavy1999,oganessian2000synthesis,oganessianSynthesisIsotopesElements2006,oganessian2010synthesis,oganessianInvestigation243Am2013}. In recent years, to open the eighth period, worldwide efforts have been made to produce SHEs beyond Oganesson (Og). Unfortunately, no correlated decay chains were observed. A key challenge is the optimal incident energy (OIE), which only depends on the theory \cite{zhuLawOptimalIncident2023b,bourzac2024heaviest}. \par 

It is worth noting that because of the plenty of experimental data, most of the theories can describe the \(^{48}\text{Ca}\)-induced fusion-evaporation reactions quite well. However, in order to synthesize SHEs beyond Og, projectiles heavier than \(^{48}\text{Ca}\) such as \(^{50}\text{Ti}\), \(^{54}\text{Cr}\), \(^{58}\text{Fe}\), and \(^{64}\text{Ni}\) are considered among the most promising candidates \cite{oganessian2024future}. Is it reliable to extrapolate theoretically from the well-investigated \(^{48}\text{Ca}\) induced reactions to those with heavier projectiles for predicting the OIEs?\par

Due to the complexity of the synthesis mechanisms of SHEs, particularly the presence of delicate ambiguities \cite{luSynthesisSuperheavyElements2016a,cap2024dipole}, the fusion process is not well understood theoretically\cite{bao2015theoretical,zagrebaev2015cross}. The fusion probability is particularly important for revealing the mechanism of synthesizing the SHEs, which is usually calculated by using diffusion models \cite{swiatecki2003fusion,liu2011calculation}, master equations \cite{wang2012theoretical,li2018predictions,bao2019possibility,li2023evaporation,dengExaminationPromisingReactions2023a,zhang2024predictions}, or empirical formulas \cite{naik2007measurement}. These theories describe the dynamics of the formation of compound nuclei from different degrees of freedom, mainly at the radial degree of freedom: distance $\vec{R}$ between the centers of the nuclei (corresponding to the elongation of a mononucleus) and the mass asymmetry degree of freedom: $\eta = \frac{A_1 - A_2}{A_1 + A_2}$ (\(A_1\) and \(A_2\) are the mass numbers of the two nuclei that make up the compound nuclei).\par

A lot of theoretical studies have been performed to investigate the synthesis of SHEs with $Z=119$ and 120. However, predictions in different theoretical models exhibit significant uncertainty and model dependence \cite{gates2024toward}. In this case, it is imperative to address several critical aspects to provide reliable theoretical predictions: (i) How to evaluate the uncertainty of the theoretical model? In Ref. \cite{fang2024bayesian}, the Bayesian uncertainty quantification method is employed to evaluate the uncertainty of the calculated ERCS in the dinuclear system model. (ii) How can we reduce the model dependence of theoretical predictions? One weak model-dependence method (the OIE law) is proposed based on the strong correlation between the OIEs, the Coulomb barrier height of side collision, and the $Q$ value  \cite{zhuLawOptimalIncident2023b}. (iii) Is it reliable to extrapolate predictions from ${}^{48}\mathrm{Ca}$ to ${}^{54}\mathrm{Cr}$? In this work, utilizing the latest experimental data of ${}^{50}\mathrm{Ti}$ and ${}^{54}\mathrm{Cr}$ projectiles introduced reactions \cite{oganessian2024future,gates2024toward},
we quantify the reliability of extrapolation from ${}^{48}\mathrm{Ca}$ to ${}^{54}\mathrm{Cr}$ within a one-dimensional model by employing the Fusion-by-Diffusion (FBD) concept, which is examined from two distinct perspectives: the radial degree of freedom and the mass asymmetry degree of freedom. \par
 
The FBD concept provides a concise physical image, in which the fusion probability is described by an analytical formula. It has been effectively applied in the investigation of the SHEs synthesis based on the radial degree of freedom for both cold and hot fusion reactions \cite{siwek2012predictions,hagino2018hot,sun2022microscopic,cap2022diffusion,cap2022fusion,cap2024dipole}. During the fusion stage, the system undergoes diffusion across a one-dimensional parabolic barrier to overcome the fusion barrier and form the compound nucleus. Also, as shown in the following, the only adjustable parameter is obtained from systematic behaviors, which will relatively reduce the uncertainty introduced by the theoretical model, making it well suited to investigate the extrapolation properties. In the present study, we adopt the FBD concept to investigate the theoretical extrapolation ability from $^{48}\mathrm{Ca}$ to heavier projectiles ($^{50}\mathrm{Ti}$/$^{54}\mathrm{Cr}$) for synthesizing the SHEs. In order to strengthen the reliability, in addition to the radial degree of freedom we also employ the FBD concept in the mass asymmetry degree of freedom.\par


$\emph{Theoretical model.}$ Theoretically, the synthesis of SHEs can be divided into three steps and the ERCS is calculated as the summation over all partial waves \textit{J}:
\begin{equation}
\begin{aligned}
\sigma_{\mathrm{ER}}(E_{\mathrm{c.m.}})=&\frac{\pi\hbar^{2}}{2\mu E_{\mathrm{c.m.}}}\sum_{J}(2J+1)T(E_{\mathrm{c.m.}},J)\\
&P_{\mathrm{fus}}(E_{\mathrm{c.m.}},J)W_{\mathrm{sur}}(E_{\mathrm{c.m.}},J),
\label{eq1}
\end{aligned}
\end{equation}
where \(E_{\mathrm{c.m.}}\) denotes the incident energy in the center-of-mass system. \(T(E_{\mathrm{c.m.}},J)\) is the transmission probability. \(P_{\mathrm{fus}}(E_{\mathrm{c.m.}},J)\) is the fusion probability and \(W_{\mathrm{sur}}(E_{\mathrm{c.m.}},J)\) denotes the survival probability.\par

In the capture process, the projectile nucleus overcomes the Coulomb potential barrier to form a dinuclear system. The penetration probability is given by the well-known Hill-Wheeler formula \cite{hill1954nuclear}.


\(P_\mathrm{fus}\) is calculated by using the FBD concept \cite{cap2011nucleus}:
\begin{equation}
\begin{aligned}
P_\mathrm{fus}(E_{\mathrm{c.m.}},J) = \frac{1}{2} (1 - \mathrm{erf} \sqrt{\frac{B_\mathrm{fus}(J)}{T}}).
\label{eq4}
\end{aligned}
\end{equation}

The $P_\mathrm{fus}(E_{\mathrm{c.m.}}, J)$ results from the angular momentum ($J$)-dependent potential energy surface of the colliding system. \(T\) is the temperature of the fusing system. \(B_\mathrm{fus}(J)\) is the barrier height of the opposing fusion. 
We study the complex fusion process in terms of the evolution of two degrees of freedom, radial and mass asymmetry degrees of freedom, respectively. Specific details are discussed in detail later.\par


The excited compound nucleus can evaporate light particles, such as neutrons, protons, and \(\alpha\) particles. We employ the Monte Carlo method to calculate the decay probabilities in each channel, the details of which are given in Ref. \cite{fang2024bayesian}.\par


\begin{figure}[t]
    \centering
    \includegraphics[width=8 cm]{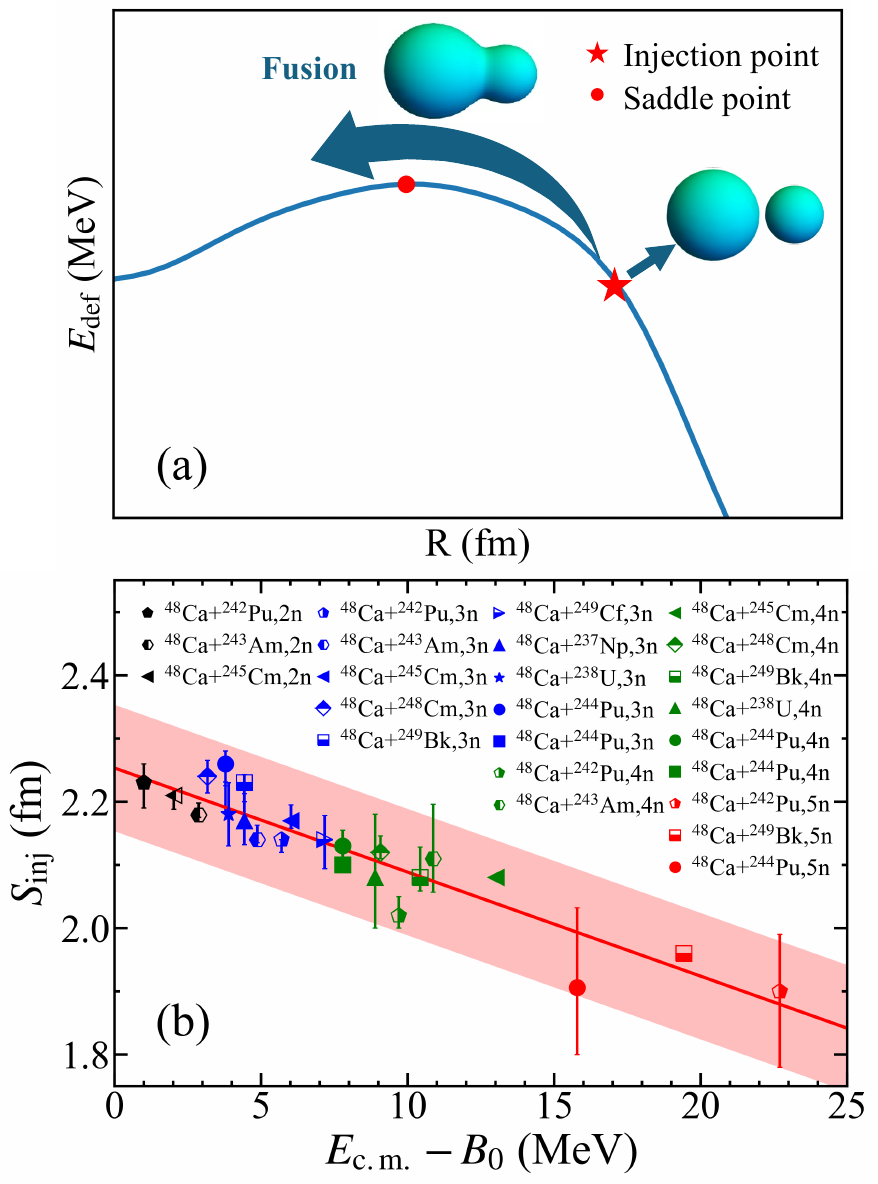}
    \caption{(Color online) 
    (a) Schematic depiction of the the fusion process in the radial degree of freedom. Red pentagram and red circle represent injection point and saddle point, respectively. 
    (b) Systematics of the $S_\mathrm{inj}$ for the hot fusion reactions with $^{48}\mathrm{Ca}$ projectiles as a function of $E_\mathrm{c.m.} - B_0$, deduced from analysis of experimental data \cite{PhysRevC.70.064609,PhysRevLett.108.022502,PhysRevC.87.014302,PhysRevLett.103.132502,PhysRevLett.105.182701,ts2000synthesis,PhysRevC.69.054607,PhysRevLett.104.252701,PhysRevC.83.054618,PhysRevC.69.021601,PhysRevC.72.034611,PhysRevC.74.044602,oganessian2000observation,oganessian2001synthesis,hofmann2012reaction,PhysRevLett.104.142502,PhysRevC.83.054315,PhysRevLett.109.162501,PhysRevC.87.054621,PhysRevLett.112.172501}.
    }
    \label{Fig.1}
\end{figure}

$\emph{Fusion at the radial degree of freedom.}$ The fusion barrier \(B_\mathrm{fus}(J)\) is a key physical quantity for calculating the fusion probability. From the view point of radial degree of freedom, the fusion could be considered as the reverse process of the fission with fixed $Q_{30}$. \(B_\mathrm{fus}(J)\) in Eq. \eqref{eq4} can be calculated as
\begin{equation}
\begin{aligned}
B_\mathrm{fus}(J) = E_\mathrm{sd} - E_\mathrm{inj} + E_\mathrm{sd}^\mathrm{rot}(J) - E_\mathrm{inj}^\mathrm{rot}(J).
\label{eq5}
\end{aligned}
\end{equation}
In this work, the deformation energies including the saddle point \(E_\mathrm{sd}\) and the injection point \(E_\mathrm{inj}\) are calculated by using the finite range liquid drop model (FRLDM) \cite{sierkMacroscopicModelRotating1986b,huangMultimodalityIr1872024}.
The distance between the surfaces of the colliding heavy-ions when the fusion begins is referred to as the injection point $S_\mathrm{inj}$.
Details of rotational energy \(E_\mathrm{sd}^\mathrm{rot}(J)\) and \(E_\mathrm{inj}^\mathrm{rot}(J)\) are given in Ref. \cite{nadtochy2021potential}.\par

\begin{figure*}[t]
    \centering
    \includegraphics[width = 0.8\textwidth]{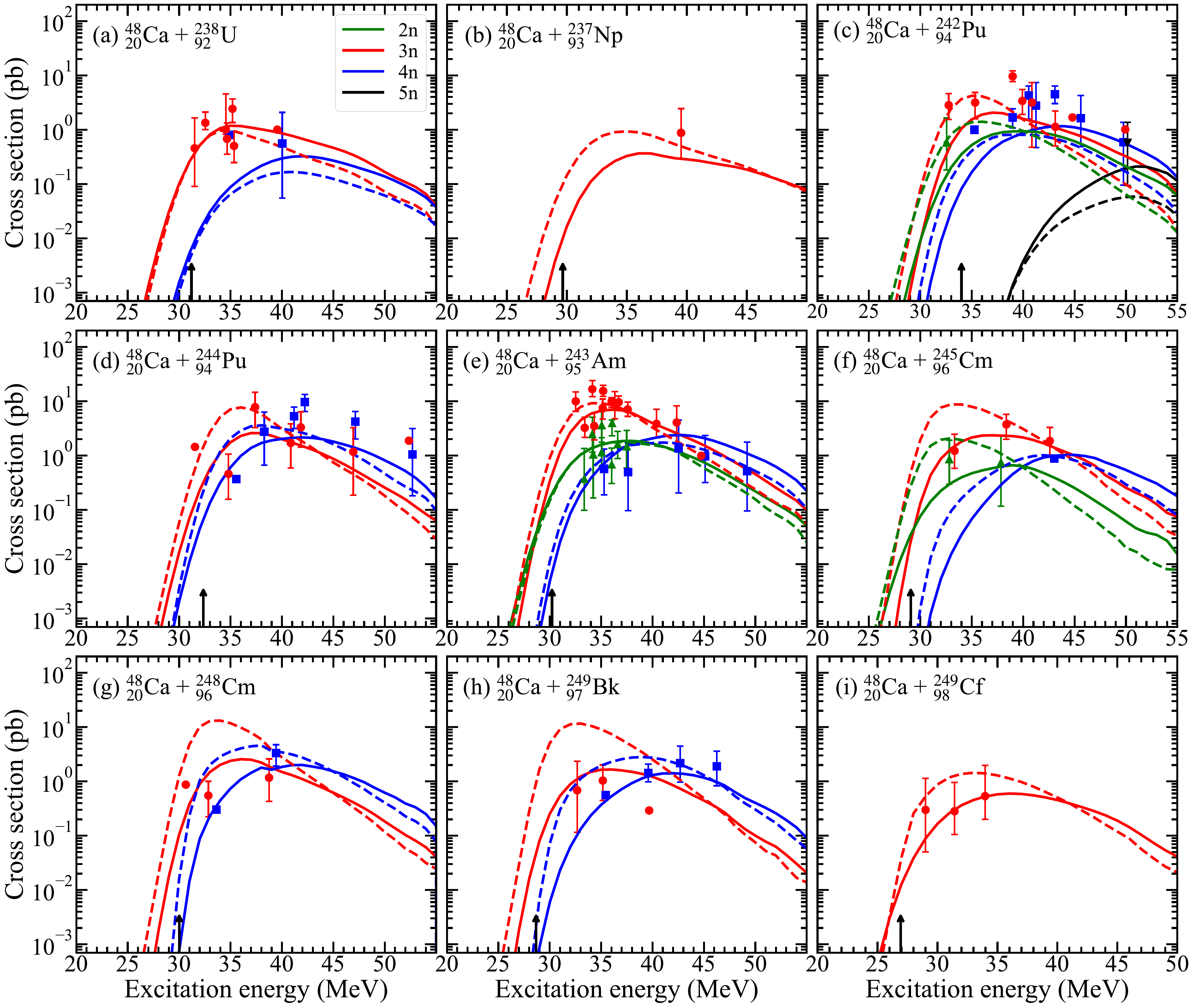}
    \caption{(Color online) In terms of radial degree of freedom (solid lines) and mass asymmetry degree of freedom (dashes lines), the calculated ERCS are compared with the available experimental data for the reactions $^{48}\mathrm{Ca}+^{238}\mathrm{U}$ \cite{PhysRevC.70.064609} (a),
    $^{48}\mathrm{Ca}+^{237}\mathrm{Np}$ \cite{PhysRevLett.108.022502,PhysRevC.87.014302} (b), $^{48}\mathrm{Ca}+^{242}\mathrm{Pu}$ \cite{PhysRevC.70.064609,PhysRevLett.103.132502,PhysRevLett.105.182701} (c), $^{48}\mathrm{Ca}+^{244}\mathrm{Pu}$ \cite{ts2000synthesis,PhysRevC.69.054607,PhysRevLett.104.252701,PhysRevC.83.054618} (d), $^{48}\mathrm{Ca}+^{243}\mathrm{Am}$ \cite{PhysRevLett.108.022502,PhysRevC.69.021601,PhysRevC.72.034611} (e), $^{48}\mathrm{Ca}+^{245}\mathrm{Cm}$ \cite{PhysRevC.69.054607,PhysRevC.74.044602,PhysRevC.87.014302} (f), $^{48}\mathrm{Ca}+^{248}\mathrm{Cm}$ \cite{oganessian2000observation,oganessian2001synthesis,PhysRevC.70.064609,hofmann2012reaction} (g), $^{48}\mathrm{Ca}+^{249}\mathrm{Bk}$ \cite{PhysRevC.87.014302,PhysRevLett.108.022502,PhysRevLett.104.142502,PhysRevC.83.054315,PhysRevLett.109.162501,PhysRevC.87.054621,PhysRevLett.112.172501} (h), and $^{48}\mathrm{Ca}+^{249}\mathrm{Cf}$ \cite{PhysRevC.74.044602,PhysRevLett.109.162501} (i). The calculated ERCS in the channels 2n, 3n, 4n, and 5n are denoted by the green lines, red lines, blue lines, and black lines, respectively. Vertical arrows denote the excitation energies corresponding to collisions at $E_\mathrm{c.m.} = B_0$ for each reaction.
    }
    \label{Fig.2}
\end{figure*}

As shown in Fig. \ref{Fig.1}(a), the fusion process is overcoming the inner fusion barrier with the evolution of the configurations from injection point to the saddle point. The position of injection point plays an important role on the fusion probability. The expression for the \(S_\mathrm{inj}\) can be derived from the experimental data by fitting the calculations to the experimentally measured maximum values of the ERCS \cite{siwek2012predictions}. The parametrization for \(S_\mathrm{inj}\) values as a function of the excess of the center-of-mass energy $E_\mathrm{c.m.}$ over the Coulomb barrier $B_0$ is shown in Fig. \ref{Fig.1}(b). The strong linear correlation between \(S_\mathrm{inj}\) and $E_\mathrm{c.m.}$ - $B_0$ is shown and the systematics can be fitted as
\begin{equation}
\begin{aligned}
    S_\mathrm{inj} = 2.253 - 0.0165 \times (E_\mathrm{c.m.} - B_0) \mathrm{fm/MeV}.
\label{eq9}
\end{aligned}
\end{equation}
The red shaded area in Fig. \ref{Fig.1}(b) indicates the error margin for \(S_\mathrm{inj}\) \cite{cap2022fusion}, which is estimated to be $\pm0.1$. The different colors of the points represent the different neutron evaporation channels. The error bars of $S_\mathrm{inj}$ are derived from the uncertainty of the experimental value.\par

It is evident from Fig. \ref{Fig.1}(b) that the injection distance $S_\mathrm{inj}$ increases as the value of $E_\mathrm{c.m.} - B_0$ decreases. This is because for the low incident energy the system has sufficient time for nucleon rearrangement, thereby facilitating neck formation at the relatively large distance. In contrast, for high incident energy, a more compressed configuration for neck formation is required. The same behavior is also shown in Ref. \cite{washiyama2008energy}. The strong correlation observed between the $S_\mathrm{inj}$ values and the corresponding energies $E_\mathrm{c.m.} - B_0$ provides support for the fission barriers proposed by Kowal et al \cite{kowal2010fission,siwek2012predictions}.\par

Eq. \eqref{eq9} determines the systematics of the injection point. Fig. \ref{Fig.2} shows the comparison of the calculated ERCS by using the Eq. \eqref{eq9} with the experimental data. 
The calculated ERCS (solid lines) can reproduce the experimental data within the error bar rather well. Therefore, the systematical behavior for $S_\mathrm{inj}$ as well as the good description of the experimental data show that the FBD concept at the radial degree of freedom is a reasonable approach for investigating the extrapolation behavior to heavier projectiles.  


\begin{figure}[t]
    \centering
    \includegraphics[width=8 cm]{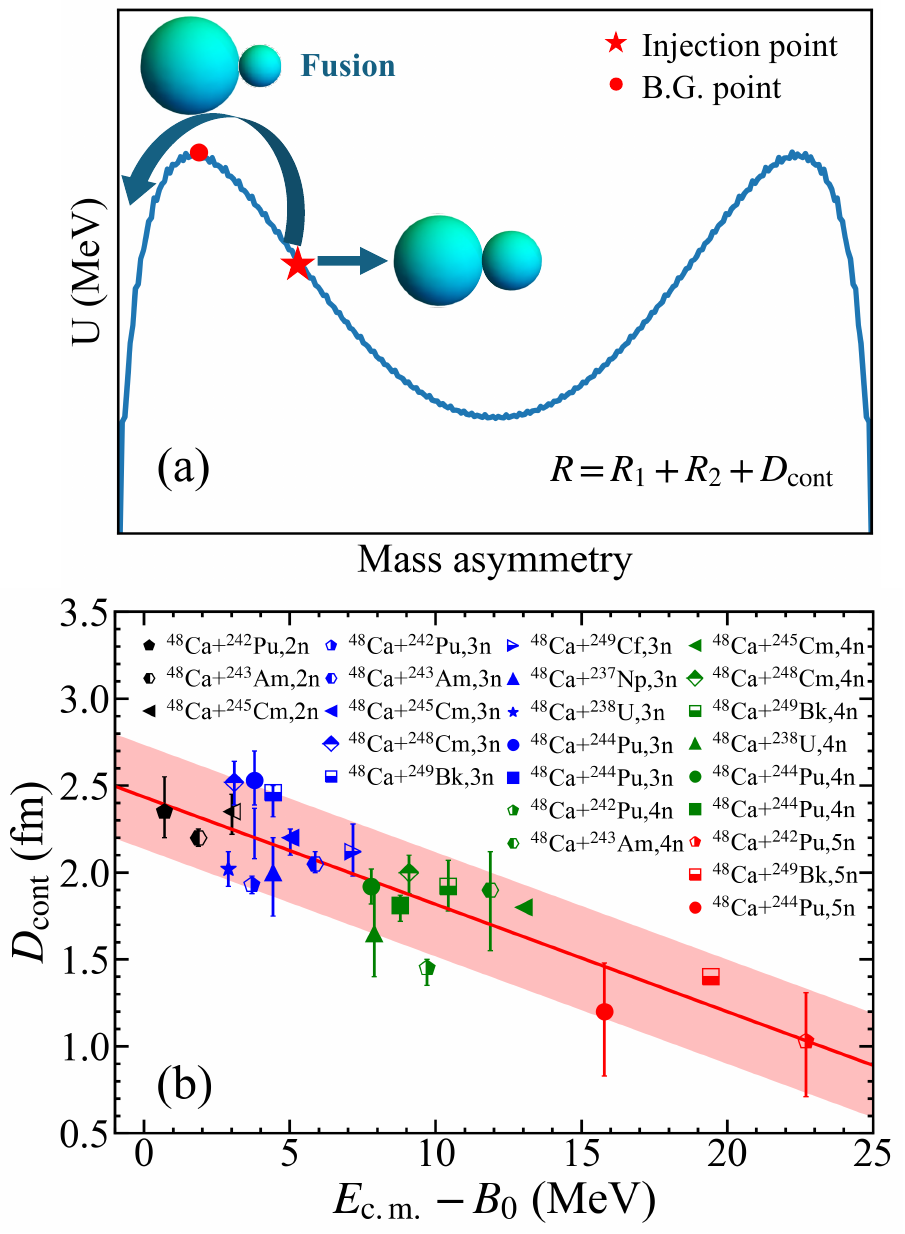}
    \caption{(Color online) 
    (a) Schematic depiction of the fusion process in mass asymmetry degree of freedom. Red pentagram and red circle represent injection point and B.G. point, respectively.
   (b) The same way as in Fig. \ref{Fig.1}(b), dependence of the $ D_\mathrm{cont}$ on $E_\mathrm{c.m.} - B_0$.
    }
    \label{Fig.3}
\end{figure}

\begin{figure*}[t]
    \centering
    \includegraphics[width=15 cm]{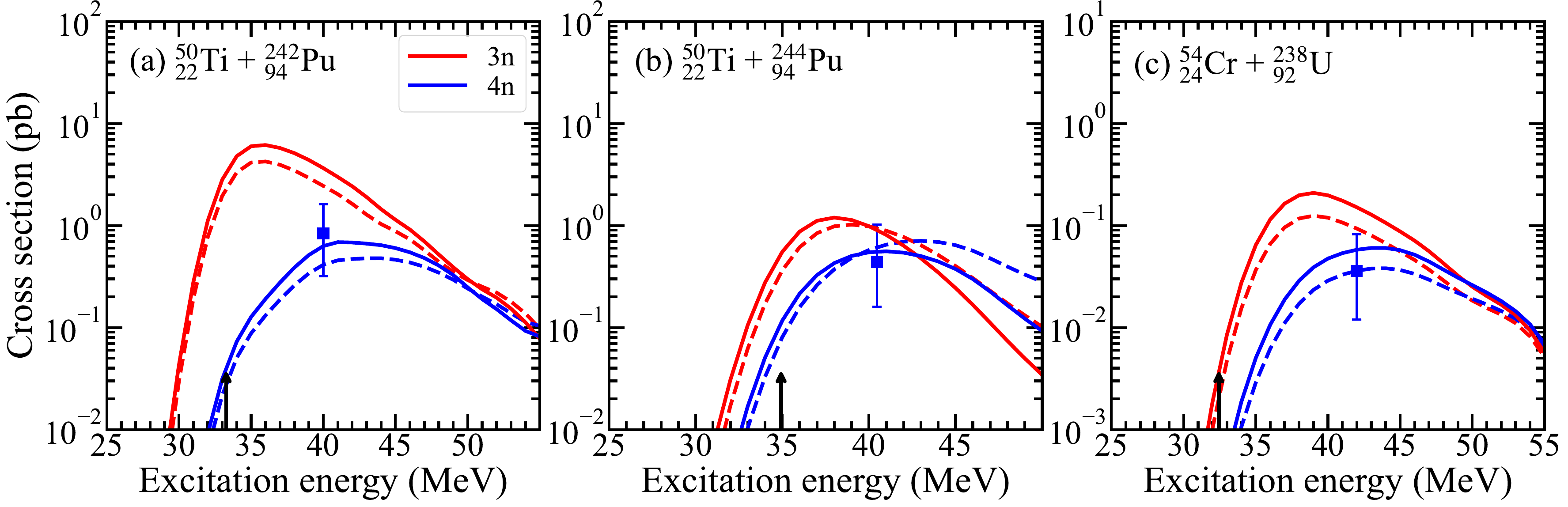}
    \caption{(Color online) Comparison of the ERCS for reactions \({}^{50}\mathrm{Ti}+{}^{242}\mathrm{Pu}\) (a),  and \({}^{50}\mathrm{Ti}+{}^{244}\mathrm{Pu}\) (b), and \({}^{54}\mathrm{Cr}+{}^{238}\mathrm{U}\) (c) in radial degree of freedom (solid lines) and mass asymmetry degree of freedom (dashed lines). Blue full squares represent experimental data for the 4n reaction channel of the reactions \({}^{50}\mathrm{Ti}+{}^{242}\mathrm{Pu}\) \cite{oganessian2024future}, \({}^{50}\mathrm{Ti}+{}^{244}\mathrm{Pu}\) \cite{gates2024toward}, and \({}^{54}\mathrm{Cr}+{}^{238}\mathrm{U}\) \cite{oganessian2024future}.}
    \label{Fig.4}
\end{figure*}

$\emph{Fusion at the mass asymmetry degree of freedom.}$ To further investigate the reliability of the FBD concept for investigating the $^{50}\text{Ti}$ and $^{54}\text{Cr}$ induced reactions for producing new SHEs, it is worth studying the fusion process from the perspective of the mass asymmetry degree of freedom. The mass asymmetry as a critical degree of freedom in the investigation the fusion probability, which has been widely studied in the dinuclear system (DNS) concept. Based on the DNS concept, in the fusion process the evolution in radial is frozen and the nucleon transfer takes place at the contact position which can be described by the distance between surfaces of the two colliding nuclei ($D_\mathrm{cont}$). The reaction of the system towards fusion generally refers to the transfer of nucleons from the lighter nucleus (either the projectile or the target) to the heavier nucleus, evolving in the direction of increasing the mass asymmetry $\eta$. The evolution of this process can be described by the diffusion of the mass asymmetry degree of freedom \(\eta\) (see Fig. \ref{Fig.3}(a)) \cite{zhang2018production}.\par

After capture, fusion takes place and the compound nucleus is formed when the DNS overcomes the inner fusion barrier \(B_\mathrm{fus}\). The more asymmetry configurations than those on the B.G. point are considered as the occurrence of fusion \cite{zhu2021unified}. \(B_\mathrm{fus}\) is calculated to be equal to the difference between the driving potential at the B.G. point and the driving potential at the injection point ($\eta$ of the projectile-target combination), as shown in Fig. \ref{Fig.3}(a). \(B_\mathrm{fus}\) is mainly determined by the details of the potential energy surface (PES) which is the potential energy of the dinuclear system along the $\eta$ direction and can be written as:
\begin{equation}
\begin{aligned}
&U(Z_1, N_1, R = R_1 + R_2 + D_\mathrm{cont}) = \Delta(Z_1, N_1) \\ 
& + \Delta(Z_2, N_2) 
+ V(Z_1, N_1, R = R_1 + R_2 + D_\mathrm{cont}),
\end{aligned}
\end{equation}
here, $R_1$ and $R_2$ are the radii of the two nuclei. $\Delta(Z_i, N_i) (i = 1, 2)$ is the mass excess of the \(i\)th fragment \cite{zhu2018shell}. $V(Z_1, N_1, R = R_1 + R_2 + D_\mathrm{cont})$ is the nucleus-nucleus interaction potential.
By assuming the rigid-body moments of inertia of different shapes, the rotational energies at the injection point and saddle point are calculated for different configurations \cite{cap2011nucleus}.\par

Similarly, the relationship between the contact distance $D_\mathrm{cont}$ and $E_\mathrm{c.m.}-B_0$ is shown in Fig. \ref{Fig.3}(b), within the error bar a linear relationship between $ D_\mathrm{cont}$ and $E_\mathrm{c.m.}-B_0$ is obtained. This interesting behavior represent that the consideration of FBD concept at the mass asymmetry degree of freedom is reasonable. The systematical behavior can be written as 
\begin{equation}
\begin{aligned}
     D_\mathrm{cont} = 2.435 \mathrm{fm} - 0.0618 \times (E_\mathrm{c.m.} - B_0) \mathrm{fm/MeV}.
\label{eq12}
\end{aligned}
\end{equation}

Unexpectedly, we observe that the inner fusion barrier exhibits a systematic similarity in the mass asymmetry and radial degrees of freedom. This suggests that to some extent the descriptions of the fusion process of both degrees of freedom share similarities and are comparable. This is why both the DNS model and the FBD model have been effectively employed to investigate fusion-evaporation reactions, with a key component of this being the accurate capture of crucial degrees of freedom. The DNS model specifically focuses on the mass asymmetry degree of freedom, while the FBD model emphasizes understanding the radial degree of freedom.\par

In order to testify the method, by using Eq. \eqref{eq12} the calculated ERCS in \(^{48}\text{Ca}\) induced reactions are compared with the experimental data, as shown in Fig. \ref{Fig.2}. The calculated results (dashed lines) are in good agreement with both the experimental ERCS and the optimal energies. The above results give us confidence based on the FBD concept at mass asymmetry degree of freedom to investigate the ERCS of fusion reactions leading to new elements.\par


$\emph{The extrapolation of projectiles with $^{50}\text{Ti}$ and $^{54}\text{Cr}$.}$ In the above calculations, we notice that the FBD concept based on the radial and mass asymmetry degrees of freedom describes the experimental data in $^{48}\text{Ca}$ induced reactions quite well. As we mentioned above, we need to clarify whether it is reasonable to extrapolate the model including the systematical behaviors of $S_\mathrm{inj}$ and $D_\mathrm{cont}$ in $^{50}\mathrm{Ti}$ and $^{54}\mathrm{Cr}$ induced reactions. Based on empirical inputs determined by the form of Eq. \eqref{eq9} and Eq. \eqref{eq12}. The calculated results in the reactions \({}^{50}\mathrm{Ti}+{}^{242}\mathrm{Pu}\), \({}^{50}\mathrm{Ti}+{}^{244}\mathrm{Pu}\), and \({}^{54}\mathrm{Cr}+{}^{238}\mathrm{U}\) are shown in Fig. \ref{Fig.4}. The calculated results for all reactions are in good agreement with the experimental data for both perspectives of radial and mass asymmetry degrees of freedom. We would like to state that the transition from $^{48}\mathrm{Ca}$ to heavier projectiles, such as $^{50}\mathrm{Ti}$ and $^{54}\mathrm{Cr}$ is reasonable and reliable according to current theories.\par

Calculations show that the ERCS for the heavier projectiles $^{50}\mathrm{Ti}$ and $^{54}\mathrm{Cr}$ in the synthesis of SHEs with $Z = 116$ are nearly one order of magnitude lower than that for the reaction induced by the $^{48}\mathrm{Ca}$ projectile. This reduction is primarily due to the lower mass asymmetry in the entrance channel of the reactions involving $^{50}\mathrm{Ti}$ and $^{54}\mathrm{Cr}$ as projectiles, coupled with a higher fusion barrier. As a result, the system has a lower probability of fusion through diffusion.\par

\begin{figure*}[t]
    \centering
    \includegraphics[width=15 cm]{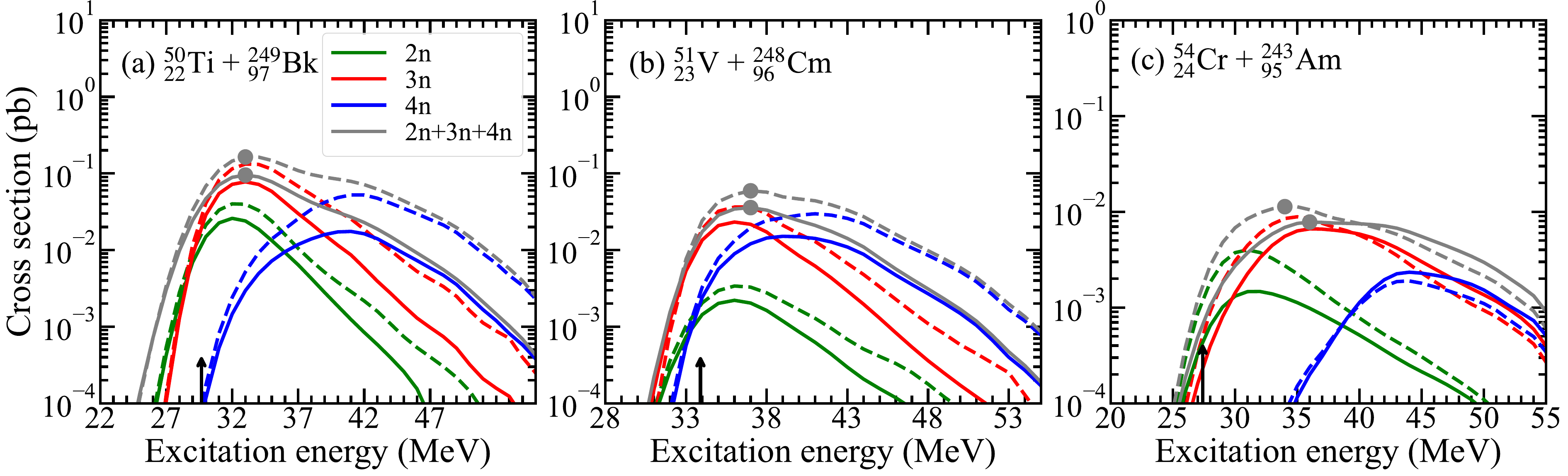}
    \caption{(Color online) 
    The ERCS production of new elements 119 through the reactions \({}^{50}\mathrm{Ti}+{}^{249}\mathrm{Bk}\) (a), \({}^{51}\mathrm{V}+{}^{248}\mathrm{Cm}\) (b), and  \({}^{54}\mathrm{Cr}+{}^{243}\mathrm{Am}\) (c) predicted in radial degree of freedom (solid lines) and mass asymmetry degree of freedom (dashed lines). The gray solid circles represent the maximum value of the ERCS for the 2n + 3n + 4n channels, corresponding to the optimal excitation energies.}
    \label{Fig.5}
\end{figure*}


$\emph{The ERCS for synthesizing the SHE with Z=119.}$ To further investigate ERCS of synthesizing SHE with Z = 119. In Fig. \ref{Fig.5}, we conducted a study on the possibility of synthesizing superheavy nuclei $Z = 119$ using $^{50}\text{Ti}$, $^{51}\text{V}$, and $^{54}\text{Cr}$ as the projectiles. In the radial degree of freedom, the OIEs for producing the element with Z = 119 via the reactions \({}^{50}\mathrm{Ti}+{}^{249}\mathrm{Bk}\), \({}^{51}\mathrm{V}+{}^{248}\mathrm{Cm}\), and \({}^{54}\mathrm{Cr}+{}^{243}\mathrm{Am}\) are estimated to be 224.5 MeV, 232.3 MeV, and 242.3 MeV, respectively. Correspondingly, under the mass asymmetric degree of freedom, the OIEs are predicted to be 224.5 MeV, 232.3 MeV, and 240.3 MeV, respectively. The predicted OIEs are close to the results from the OIE law proposed in Ref \cite{zhuLawOptimalIncident2023b}, especially for the reactions \({}^{50}\mathrm{Ti}+{}^{249}\mathrm{Bk}\) and \({}^{54}\mathrm{Cr}+{}^{243}\mathrm{Am}\).\par


$\emph{Conclusions.}$ The ERCS for synthesizing SHEs based on the different perspectives (radial and mass asymmetry degrees of freedom) within the FBD concept are investigated. By calibrating the injection point distance $S_\mathrm{inj}$ and contact distance $D_\mathrm{cont}$ as adjustable parameters against the experimental ERCS data from the \(^{48}\text{Ca}\)-induced fusion reactions in the different perspectives. We observe the strong linear systematic behaviors of the parameters on $E_\mathrm{c.m.}-B_0$. The inner fusion barrier shows a systematic similarity in both the radial and mass asymmetry degrees of freedom, which suggests that to some certain extent radial and mass asymmetry degrees of freedom are similar and comparable in their description of the fusion process. The parameterization of $S_\mathrm{inj}$ and $D_\mathrm{cont}$ are then used for extrapolation to calculate the ERCS in the reactions \({}^{50}\mathrm{Ti}+{}^{242}\mathrm{Pu}\), \({}^{50}\mathrm{Ti}+{}^{244}\mathrm{Pu}\), and \({}^{54}\mathrm{Cr}+{}^{238}\mathrm{U}\). The calculated results show good agreement with the recent experimental data from LBNL \cite{gates2024toward} and Dubna \cite{oganessian2024future}, which indicate that the theoretical calculations are relatively reliable in extrapolating the projectiles from $^{48}\text{Ca}$ to $^{50}\text{Ti}$ and $^{54}\text{Cr}$ for synthesizing SHEs beyond Og. Finally, we present predictions of ERCS and OIEs for the synthesis of SHN with $Z = 119$ from both radial and mass asymmetry degrees of freedom. The predicted OIEs in radial and mass asymmetry degrees of freedom are consistent with each other, as well as the results in Ref. \cite{zhuLawOptimalIncident2023b}. 
This work also inspires us to utilize microscopic theories (such as density functional theories \cite{benderPotentialEnergySurfaces1998,sekizawaTimedependentHartreeFockLangevin2019a,sun2022microscopic} for calculating multidimensional PES and address it in the FBD framework.


$\emph{Acknowledgments.}$ The authors would like to thank Ying-Ge Huang and Fu-Chang Gu for the FRLDM calculations; to Feng-Shou Zhang, Shan-Gui Zhou, Zai-Guo Gan, Cheng-Jian Lin, Zhong-Zhou Ren, Hong-Fei Zhang, Xiao-Jun Bao, Ning Wang, Nan Wang, Jun-Chen Pei, Zhi-Yuan Zhang, Xiao-Tao He, Hui-Min Jia, Hua-Bin Yang, Bing Wang, Jing-Jing Li, and Gen Zhang for useful discussions. This work was supported by the National Natural Science Foundation of China under Grants No. 12075327 and 12475136; The Open Project of Guangxi Key Laboratory of Nuclear Physics and Nuclear Technology
under Grant No. NLK2022-01; Fundamental Research Funds for the Central Universities, Sun Yat-sen University under Grant No. 23lgbj003.

\bibliographystyle{ieeetr} 
\bibliography{bibliography}

\end{document}